\documentstyle[floats,prd,aps,epsf]{revtex}
\draft 

\begin{document}

\twocolumn[\hsize\textwidth\columnwidth\hsize\csname
@twocolumnfalse\endcsname

\title{General Relativistic Models of Binary Neutron Stars 
	in Quasiequilibrium}

\author{T.~W.~Baumgarte}
\address{Department of Physics, University of Illinois at
        Urbana-Champaign, Urbana, Il~61801}
\author{G.~B.~Cook, M.~A.~Scheel}
\address{Center for Radiophysics and Space Research, Cornell University,
        Ithaca, NY 14853}
\author{S.~L.~Shapiro}
\address{Departments of Physics and Astronomy and
	National Center for Supercomputing Applications, 
	University of Illinois at Urbana-Champaign, Urbana, Il~61801}
\author{S.~A.~Teukolsky}
\address{Departments of Physics and Astronomy and
	Center for Radiophysics and Space Research, Cornell University,
        Ithaca, NY 14853}
\maketitle

\begin{abstract}
We perform fully relativistic calculations of binary neutron
stars in corotating, circular orbit.  While Newtonian gravity allows
for a strict equilibrium, a relativistic binary system emits
gravitational radiation, causing the system to lose energy and slowly
spiral inwards.  However, since inspiral occurs on a time scale much
longer than the orbital period, we can treat the binary to be in
quasiequilibrium. In this approximation, we integrate a subset of the
Einstein equations coupled to the relativistic equation of hydrostatic
equilibrium to solve the initial value problem for binaries of
arbitrary separation. We adopt a polytropic equation of state to
determine the structure and maximum mass of neutron stars in close
binaries for polytropic indices $n=1$, 1.5 and 2.  We construct
sequences of constant rest-mass and locate turning points along
energy equilibrium curves to identify the onset of orbital
instability. In particular, we locate the innermost stable circular
orbit (ISCO) and its angular velocity. We construct the first contact
binary systems in full general relativity. These arise whenever the
equation of state is sufficiently soft ($n \stackrel{>}{\sim} 1.5$).
A radial stability analysis reveals no tendency for neutron stars
in close binaries to collapse to black holes prior to merger.
\end{abstract}

\pacs{PACS numbers: 04.20.Ex, 04.25.Dm, 04.30.Db, 04.40.Dg, 97.60.Jd}

\vskip2pc]


\section{INTRODUCTION}

Neutron star binaries are interesting for numerous reasons. Several
neutron star binary systems are known to exist even within our own
galaxy~\cite{tamt93}. For some of these systems (including PSR
B1913+16, B1534+12) general relativistic effects in the binary orbit
have been measured to high precision~\cite{tw89,a95}.  Binary neutron
stars are believed to be among the most promising sources of
gravitational waves for detectors like LIGO, VIRGO and GEO. This
circumstance has triggered multiple efforts to predict the
gravitational waveform emitted during the inspiral and the final
plunge of the two stars.  More fundamentally, the two-body problem is
one of the outstanding unsolved problems in classical general
relativity.

Considerable effort has gone into understanding binary
neutron stars. Most of this work has been performed within the
framework of Newtonian hydrodynamics. Hachisu and Eriguchi~\cite{he84}
constructed hydrostatic equilibrium of binaries in synchronized
circular orbits. Rasio and Shapiro~\cite{rs92} studied binary
equilibrium configurations and their dynamical evolution, including
the merger of the two stars.  The coalescence of neutron star binaries
has also been investigated by Shibata, Nakamura and
Oohara~\cite{sno92}, Zhuge, Centrella and McMillan~\cite{zcm94} and
Ruffert, Janka and Sch\"afer~\cite{rjs95} and other investigators.

Many investigators have also studied the binary problem within a
post-Newtonian framework. As long as the stars are well separated they
can be approximated by point sources. In this case hydrodynamical
effects are neglected and the gravitational waveform can be 
calculated to second post-Newtonian order (see~\cite{bdiww95} and
references therein). Post-Newtonian calculations that do take into
account hydrodynamical effects are also under way: Shibata~\cite{s96}
and Taniguchi and Shibata~\cite{ts97} have constructed equilibrium
configurations and Oohara and Nakamura~\cite{on96} have studied binary
coalescence.  Lai~\cite{l96}, Lai and Wiseman~\cite{lw96} and
Lombardi, Rasio and Shapiro~\cite{lrs97} have constructed binary
equilibrium configurations in an ellipsoidal approximation.

Fully general relativistic treatments of the problem are complicated
by several factors, including the non-linearity of the partial
differential equations and the requirement of very large computational
resources to solve the coupled system.  These simulations are
currently only in their infancy~\cite{on96}.  Recently, Wilson,
Mathews and Marronetti~\cite{wm95} (hereafter WMM) reported results
obtained with a relativistic numerical code. Their code
assumed several simplifying physical and mathematical approximations.
Their results suggest that the central densities of the stars increase
as the stars approach each other and that massive neutron stars
individually collapse to black holes prior to merger. WMM therefore
find that in general relativity, the presence of a companion star and
its tidal field tend to destabilize the stars in a binary system. This
conclusion is opposite to what is expected from
Newtonian~\cite{lrs93}, post-Newtonian~\cite{l96,lw96,lrs97,w97},
perturbative~\cite{bh97} and matched asymptotic
expansion~\cite{f97,t97} treatments of the problem. WMM also find that
just prior to plunge and merger, their binary system has a total
angular momentum too large to form a Kerr black hole (see the
discussion in~\cite{eh96}).

In this paper we construct fully relativistic binary neutron stars in
quasiequilibrium circular orbit (``quasi''-equilibrium because these
binaries are not strictly stationary: because of the slow emission of
gravitational radiation, general relativistic binaries cannot be in
strict equilibrium).  These models are interesting on their own right
and provide initial data for future dynamical evolution
calculations. We study the structure of the neutron stars in these
close binary systems and determine, for example, their maximum allowed
equilibrium mass.  In addition, we build quasiequilibrium binary
sequences of constant rest-mass. These sequences approximate evolutionary
trajectories of neutron star binaries undergoing slow inspiral via
the generation of gravitational radiation. By locating the turning
points in their total energy versus separation curves, we can identify
the onset of orbital instability at the innermost stable circular
orbit (ISCO) and the orbital parameters at that critical radius.  We
have presented preliminary results in~\cite{bcsst97a}, and analyzed
the stability of these binaries in~\cite{bcsst97b}. We do not find any
evidence for a destabilization of neutron stars in close binaries.

The purpose of this paper is to discuss details of our approximations,
equations, and numerical method, and to present more complete results.
The paper is organized as follows: in Section~\ref{assumptions} we
discuss all the underlying assumptions and approximations made in our
calculations. In Section~\ref{equations} we derive all the equations
describing the quasiequilibrium of relativistic binary neutron stars.
The numerical implementation of these equations is described in
Section~\ref{numerics}. We present results for several different
polytropic equations of state in Section~\ref{results} and briefly
summarize our findings in Section~\ref{summary}. We also include an
Appendix with tabulated data for some of our sequences.


\section{BASIC ASSUMPTIONS AND APPROXIMATIONS}
\label{assumptions}

Throughout this paper we will assume that the two neutron stars have
equal mass, are corotating in a circular orbit and that the matter
obeys a polytropic equation of state.

Choosing a polytropic equation of state permits a wide survey of
models as a function of the stiffness of the equation of state and
also simplifies the integration of the matter
equation~(\ref{enthalpy0}).  However, polytropic equations could be
easily replaced by more realistic cold equations of state.

Restricting our analysis to
stars with equal masses allows us to exploit spatial
symmetry and solve the problem in just one octant in our Cartesian
grid (see
Section~\ref{fieldequ} below). However, generalizing our method to
stars of unequal mass is straightforward.  Nevertheless, it is
interesting to note that all well determined masses of neutron stars in
close binary systems have masses remarkably close to $1.4 M_{\odot}$
(see, for example,~\cite{tamt93}).  Focussing on stars with equal
mass may therefore be physically reasonable as well as numerically
convenient.

Demanding that the stars be corotating is a much less realistic
assumption. Even if the stars in a binary started out corotating at a
large separation, maintaining this corotation during inspiral would
require a larger viscosity than is possible in neutron
stars~\cite{bc92,k92}.  Instead, it is more likely that the
circulation of the stars is conserved during inspiral.  However, our
assumption of corotation greatly simplifies the solution of the
problem (see Section~\ref{matterequ}) and it is appropriate to tackle
this simpler case first. Even in Newtonian theory, the construction of
nonsynchronous binaries is difficult because of the unknown velocity
field; only in ellipsoidal models can one build nonsynchronous as
easily as synchronous binaries~\cite{lrs97}.  Constructing more
realistic sequences of constant circulation requires a dynamical
treatment, as one marches inward from one radius to the next using the
full coupled set of field and hydrodynamic evolution equations to
guarantee conservation of circulation.

In Newtonian gravity, a strict equilibrium solution for two such stars
in a synchronized circular orbit always exists, except for very stiff
equations of state (with $n \stackrel{<}{\sim} 1.5$) near
contact~\cite{he84}.  Since this solution is stationary, the
hydrodynamical equations for the matter reduce to a single Bernoulli
integral, which greatly simplifies the problem (see
Section~\ref{newton}).

Because of the emission of gravitational waves, a binary in general
relativity cannot be in strict equilibrium. However, up to the
innermost stable circular orbit (ISCO), the timescale for orbital decay
by radiation will be much longer than the orbital period, so that the
binary can be considered to be in ``quasiequilibrium''. This allows
us to neglect both gravitational waves and wave-induced deviations
from a circular orbit to good approximation. A similar approximation
is often used in stellar evolution calculations: there the relevant
evolution timescales are the nuclear or Kelvin-Helmholtz timescales,
while the stars maintain (quasi) hydrostatic equilibrium on a
dynamical timescale.

We attempt to minimize the gravitational wave content by choosing the
spatial metric to be conformally flat, as in WMM (see
also~\cite{wm89}). As will be shown in Section~\ref{fieldequ}, the
field equations then reduce to a set of coupled, quasi-linear elliptic
equations for the lapse, the shift and the conformal
factor. If we neglect small deviations from circular orbit, the fluid
flow is again stationary, and the hydrodynamical equations again
reduce to a relativistic Bernoulli integral (see
Section~\ref{matterequ}). Solving these equations yields a valid
solution to the initial value (constraint) equations, and an
approximate solution to the full Einstein equations for an inspiraling
binary at any given moment, prior to plunge.

This conformal approximation has been carefully tested in
Ref.~\cite{cst96} for a single rotating star in stationary
equilibrium, which is the simplest numerical example in relativity for
which the equilibrium solution deviates from conformal flatness. In
Ref.~\cite{cst96} it was shown that by assuming conformal flatness,
the resulting deviations from the exact solution were typically much
smaller than 1\%, even for highly relativistic stars.


\section{BASIC EQUATIONS}
\label{equations}

\subsection{Field Equations}
\label{fieldequ}

To construct a numerical model of a binary system we employ the ADM
decomposition of Einstein's equations of general relativity~\cite{adm62}.
The derivation of our adopted equations closely follows the derivation 
in~\cite{cst96} for rotating stars.

We write the metric in the general form
\begin{equation}
ds^2 = - \alpha^2 dt^2 + \gamma_{ij}(dx^i - \omega^i dt)
	(dx^j - \omega^j dt).
\end{equation}
Throughout the paper Latin indices will run from 1 to 3, whereas Greek
indices will run from 0 to 3. We also set $G = c = 1$.
By definition of the extrinsic curvature $K_{ij}$, the three-metric 
$\gamma_{ij}$ satisfies the dynamical equation
\begin{equation}
\partial_{t} \gamma_{ij} = - 2 \alpha K_{ij} - D_i \omega_j - D_j \omega_i,
\end{equation}
where $D_i$ denotes the covariant derivative associated with $\gamma_{ij}$.
This equation can be decomposed into its trace
\begin{equation}
\partial_t \ln \gamma^{1/2} = - \alpha K - D_i \omega^i,
\end{equation}
where $\gamma = \det \gamma_{ij}$ and $K = K^i_{~i}$, and its trace-free
part
\begin{eqnarray} \label{tracefree}
\gamma^{1/3} \partial_t (\gamma^{-1/3} \gamma_{ij}) & = &
	- 2 \alpha (K_{ij} - \frac{1}{3} \gamma_{ij} K)  \nonumber \\
& &	- D_i \omega_j - D_j \omega_i + \frac{2}{3} \gamma_{ij} D_k \omega_k.
\end{eqnarray} 
In the following we will choose maximal slicing so that
\begin{equation} \label{maxslice}
K = 0.
\end{equation}

We expect the gravitational wave content of the spacetime to be small
(see Section~\ref{assumptions}), and we now want to use this
expectation to simplify the problem. Unfortunately, the physical
fields cannot be cleanly separated into freely specifiable dynamical
degrees of freedom and dependent quantities, which are determined by
the constraint equations. However, such an identification is possible
with the help of a conformal decomposition~\cite{y72}.  We can
therefore attempt to minimize the gravitational wave content of the
(physical) spacetime by removing the dynamical (or ``wave'') degrees of
freedom from the conformal fields. This can be achieved by choosing the
three-metric $\gamma_{ij}$ to be conformally flat, so that
$\gamma^{-1/3} \gamma_{ij} = f_{ij}$, where $f_{ij}$ is the flat space
metric.  We will later use Cartesian coordinates, for which $f_{ij}$
becomes the Kronecker delta $\delta_{ij}$. Note that this choice can
always be made to find initial data on one time slice without any
approximation.  Our approximation lies in assuming that the metric
will {\em remain} conformally flat for all
times during the inspiral. Eq.~(\ref{tracefree}) then reduces to~\cite{fn1}
\begin{equation} \label{k1}
2 \alpha K_{ij} = - D_i \omega_j - D_j \omega_i 
	+ \frac{2}{3} \gamma_{ij} D_k \omega^k.
\end{equation}

We now write the metric as
\begin{equation} \label{conflatmet}
\gamma_{ij} = \Psi^4 f_{ij},
\end{equation}
where $\Psi$ is the conformal factor. The later is determined by the 
Hamiltonian constraint
\begin{equation} \label{ham1}
R - K_{ij} K^{ij} = 16 \pi \rho,
\end{equation}
where the source term $\rho$ is defined by
\begin{equation} \label{rho}
\rho = n^\alpha n^\beta T_{\alpha\beta}.
\end{equation}
Here $n^{\alpha}$ is the normal vector to a $t = {\rm const}$ slice and
$T_{\alpha\beta}$ is the stress-energy tensor.  For the
metric~(\ref{conflatmet}), the Ricci scalar $R$ in~(\ref{ham1}) reduces to
\begin{equation}
R = - 8 \Psi^{-5} \nabla^2 \Psi,
\end{equation}
where $\nabla^2$ is the flat space Laplacian associated with $f_{ij}$.
Inserting this into~(\ref{ham1}) we find
\begin{equation} \label{ham2}
\nabla^2 \Psi = - \frac{1}{8} \Psi^{-7} \tilde K_{ij} \tilde K^{ij}
	- 2\pi\Psi^5 \rho.
\end{equation}
Here we have transformed $K^{ij}$ according to
\begin{equation}
\tilde K^{ij} = \Psi^{10} K^{ij},
\end{equation}
which, from Eq.~(\ref{k1}), now satisfies
\begin{equation} \label{k2}
\tilde K^{ij} = - \frac{\Psi^6}{2\alpha} 
	\left( \nabla^i \omega^j + \nabla^j \omega^i
	- \frac{2}{3} f^{ij} \nabla_k \omega^k \right).
\end{equation}
Inserting this expression into the momentum constraint
\begin{equation} \label{mom1}
D_j K^{ij} = 8 \pi j^{i}
\end{equation}
yields
\begin{equation} \label{mom2}
\nabla^2 \omega^i + \frac{1}{3} \nabla^i (\nabla_j \omega^j) 
= 2 \nabla_j \ln( \alpha \Psi^{-6} ) \tilde K^{ij} - 16 \pi \alpha \Psi^4 j^i.
\end{equation}
Here the source term $j^i$ is given by
\begin{equation} \label{j}
j^{\alpha} = - \gamma^{\alpha}_{~\beta}n_{\gamma} T^{\beta\gamma}.
\end{equation}
This equation can be simplified by writing the shift vector as a sum 
of a vector and a gradient~\cite{by80}
\begin{equation} \label{omega}
\omega^i = G^i - \frac{1}{4} \nabla^i B.
\end{equation}
Eq.~(\ref{mom2}) can then be replaced by the two equations
\begin{equation} \label{g1}
\nabla^2 G^i   
= 2 \nabla_j \ln( \alpha \Psi^{-6} ) \tilde K^{ij} - 16 \pi \alpha \Psi^4 j^i
\end{equation}
and 
\begin{equation} \label{b1}
\nabla^2 B = \nabla_i G^i.
\end{equation}

Imposing the full set of dynamical equations for the evolution of
$K_{ij}$ would be inconsistent with Eq.~(\ref{k1}) and our approximation
that $\gamma_{ij}$ remains conformally flat at all times. However, in
addition to Eq.~(\ref{maxslice}) we can always require that the maximal
slicing condition be preserved $\partial_t K = 0$. Taking the trace of
the time evolution equation for $K_{ij}$ together with
Eq.~(\ref{ham2}) then yields an equation for the lapse
\begin{equation} \label{lap2}
\nabla^2(\alpha \Psi) = \alpha \Psi \left( \frac{7}{8} \Psi^{-8} 
	\tilde K_{ij} \tilde K^{ij}  + 2 \pi \Psi^4 (\rho + S) \right).
\end{equation}
Here the source term $S$ is defined by
\begin{equation} \label{S}
S = \gamma^{ij} T_{ij}.
\end{equation}

Eqs.~(\ref{ham2}), (\ref{g1}), (\ref{b1}) and (\ref{lap2}) together
with the matter equations (see next Section) form a system of coupled,
nonlinear elliptic equations, which has to be solved iteratively. The
boundary conditions follow from asymptotic flatness.  Following
Bowen~\cite{b82}, the exterior solution to the field equations can be
expanded in terms of multipole moments. We adopt as outer boundary
conditions the fall-off behavior of the lowest order non-vanishing
multipole moments.  Because of the symmetries of the problem it is
possible to solve it in only one octant of a Cartesian grid. The
resulting boundary conditions on the coordinate planes together with
the outer boundary conditions are summarized in Table~1.

\begin{table}
\begin{center}
\begin{tabular}{cccc}
$r \rightarrow \infty$ & $x = 0$ & $y = 0$ & $z = 0$ \\
\tableline
$G^x \sim \displaystyle \frac{z}{r^3}$ & 
$\partial_x G^x = 0$	& $\partial_y G^x = 0$ &  $G^x = 0$ \\[3mm]
$G^y \sim  \displaystyle \frac{xyz}{r^7}$ &
$G^y = 0$		& $G^y = 0$		& $G^y = 0$ \\[3mm]
$G^z \sim \displaystyle \frac{x}{r^3}$ &
$G^z = 0$		& $\partial_y G^z = 0$	& $\partial_z G^z = 0$\\[3mm]
$B \sim \displaystyle \frac{xz}{r^3}$ &
$B = 0$			& $\partial_y B$ = 0	& $B = 0$\\[3mm]
$\alpha - 1 \sim \displaystyle \frac{1}{r}$ &
$\partial_x \alpha = 0$	& $\partial_y\alpha=0$&$\partial_z\alpha = 0$\\[3mm] 
$\Psi - 1 \sim \displaystyle \frac{1}{r}$ &
$\partial_x \Psi = 0$	& $\partial_y \Psi = 0$	& $\partial_z \Psi = 0$\\
\end{tabular}
\end{center}
\caption{Boundary conditions for the outer boundaries ($r \rightarrow \infty$)
and on the coordinate planes in Cartesian coordinates. The equatorial 
plane is taken to be the $y = 0$ plane and the stars are taken to be 
aligned with the $z$-axis.}
\end{table}


\subsection{Matter Equations}
\label{matterequ}

As we have discussed in Section~\ref{assumptions}, we neglect 
wave-induced deviations from a strictly periodic, circular orbit, and 
also assume the stars to be corotating.
In Cartesian coordinates we can choose the equatorial plane to be the
$y=0$ plane, so that the fluid four velocity then takes the form
\begin{equation}
u^\alpha = u^t(1,\Omega z,0, -\Omega x),
\end{equation}
where $\Omega$ is the constant angular velocity. We introduce a vector
\begin{equation} \label{xi}
\xi^\alpha = (0,z,0,-x),
\end{equation}
in terms of which the four velocity can also be written
\begin{equation}
u^\alpha = u^t (\alpha n^\alpha + \Omega\xi^\alpha - \omega^\alpha).
\end{equation}
Define $v$ to be the relative velocity between the matter
and a normal observer
\begin{equation} \label{ut}
\frac{1}{(1 - v^2)^{1/2}} = - n_{\alpha}u^{\alpha} = \alpha u^t.
\end{equation}
Then, from $u^\alpha u_\alpha = -1$, we find
\begin{equation}
v^2 = \frac{\Psi^4}{\alpha^2} \left( (\Omega z - \omega^x)^2 +
	(\omega^y)^2 + (\Omega x + \omega^z)^2 \right).
\end{equation}

For a perfect fluid the stress energy tensor is
\begin{equation}
T^{\alpha\beta} = (\rho_0 + \rho_i + P) u^\alpha u^\beta + P g^{\alpha\beta},
\end{equation}
where $\rho_0$ is the rest-mass density, $\rho_i$ is the internal energy
density and $P$ is the pressure. The source term $\rho$ 
in Eq.~(\ref{rho}) can then be written
\begin{equation} \label{rho2}
\rho 	= \frac{\rho_0 + \rho_i + P}{1 - v^2} - P,
\end{equation}
the momentum source $j^i$ in Eq.~(\ref{j}) becomes
\begin{equation} \label{j2}
j^i 	= \frac{(\rho_0 + \rho_i + P)}{\alpha} 
	\frac{(\Omega \xi^i - \omega^i)}{1-v^2},
\end{equation}
and $S$ in Eq.~(\ref{S}) is given by
\begin{equation} \label{S2}
S 	= (\rho_0 + \rho_i + P)\frac{v^2}{1 - v^2} + 3P.
\end{equation}

In order to describe the matter close to equilibrium we will use
two of our basic assumptions. Neglecting
deviations from a strictly periodic circular orbit and taking
the two stars to be corotating is equivalent to assuming that the
fluid four velocity is proportional to a Killing vector
\begin{equation}
\frac{\partial}{\partial t} + \Omega \frac{\partial}{\partial \phi}.
\end{equation}
In this approximation, the matter equations can be integrated
analytically, which yields the relativistic Bernoulli integral
(see, e.g.,~\cite{lppt75})
\begin{equation} \label{eul1}
\frac{u^t}{h} = {\rm const}.
\end{equation}
Here $h$ is the enthalpy
\begin{equation} \label{enthalpy0}
h = \exp \left( \int \frac{dP}{\rho_0 + \rho_i + P} \right).
\end{equation}
For a polytropic equation of state
\begin{equation} \label{eos}
P = \kappa \rho_0^{1 + 1/n},
\end{equation}
where $\kappa$ is the polytropic constant and $n$ the polytropic index, the
enthalpy becomes
\begin{equation} \label{enthalpy}
h = \frac{\rho_0 + \rho_i + P}{\rho_0}.
\end{equation}

It is very useful to introduce a dimensionless ratio
\begin{equation}
q = \frac{P}{\rho_0},
\end{equation}
in terms of which we can express
\begin{eqnarray}
\rho_0 & = & \kappa^{-n} q^{n} \label{rho_0} \\[2mm]
\rho_i & = & n \kappa^{-n} q^{n+1} \\[2mm] 
P & = & \kappa^{-n} q^{n+1}. \label{P}
\end{eqnarray}
Note that in the Newtonian limit we have $q \ll 1$.  Inserting the
last three expressions together with Eqs.~(\ref{ut}) and~(\ref{enthalpy})
into Eq.~(\ref{eul1}) we find
\begin{equation} \label{eul3}
q = \frac{1}{1+n} \left( \frac{1 + C}{\alpha (1 - v^2)^{1/2}} -1 \right),
\end{equation}
where we have written the constant in Eq.~(\ref{eul1}) as $1+C$. Also,
we use $q$ to rewrite the source terms~(\ref{rho2}--\ref{S2}) as
\begin{eqnarray} \label{source}
\rho & = & \displaystyle \kappa^{-n} q^n \left( \frac{1 + (1+n)q}{1 - v^2} - q \right) \\[2mm]
j^i & = & \displaystyle \kappa^{-n} q^n \frac{(1 + (1+n)q)}{\alpha}
	\frac{(\Omega \xi^i - \omega^i)}{1 - v^2} \label{jsource} \\[2mm]
\rho + 2S & = & 
	\displaystyle \kappa^{-n}q^n \left( \frac{1 + (1+n)q}{1 - v^2}(1+2v^2) + 5q \right).
\end{eqnarray}

Note that physical dimensions enter our problem only through the
polytropic constant $\kappa$ in the equation of state~(\ref{eos}). It
is therefore useful to nondimensionalize all equations and eliminate
$\kappa$ from the problem. This means that given the polytropic index
$n$, we can solve the equations once and use the results for arbitrary
$\kappa$.  Since $\kappa^{n/2}$ has units of length we can introduce
dimensionless coordinates $\bar t = \kappa^{-n/2}t$, $\bar x =
\kappa^{-n/2}x$ and the same for $y$ and $z$. The derivative operator
scales as $\bar \nabla_i = \kappa^{n/2}
\nabla_i$, and the extrinsic curvature as $\bar K^{ij} =
\kappa^{n/2} \tilde K^{ij}$.  
The angular velocity $\Omega$ transforms according
to $\bar \Omega = \kappa^{n/2} \Omega$. We also rescale
$\bar B = \kappa^{-n/2} B$ and $\bar \xi^i = \kappa^{-n/2} \xi^i$
Putting terms
together we find the Hamiltonian constraint
\begin{equation} \label{ham3}
\bar \nabla^2 \Psi = -\frac{1}{8} \Psi^{-7} \bar K_{ij} \bar K^{ij}
	- 2 \pi \Psi^5 q^n \left( \frac{1+(1+n)q}{1-v^2} - q \right),
\end{equation}
the lapse equation
\begin{eqnarray} \label{lap3}
\bar \nabla^2 \tilde \alpha & = & \tilde \alpha \frac{7}{8} \Psi^{-8}
 	\bar K_{ij} \bar K^{ij} \nonumber \\
	& & + 2 \pi  \tilde \alpha \Psi^4 q^n 
	\left( (1+(n+1)q) \frac{1+2v^2}{1-v^2} + 5q \right),
\end{eqnarray}
and the momentum constraint equations
\begin{eqnarray} \label{g3}
\bar \nabla^2 G^i & = & 
	-2 \bar \nabla_j (\tilde \alpha \Psi^{-7}) \bar K^{ij} 
	\nonumber \\
	& & -16 \pi \Psi^4 q^n \frac{1+(1+n)q}{1-v^2}
	(\bar \Omega \bar \xi^i - \omega^i)
\end{eqnarray}
and
\begin{equation} \label{b3}
\bar \nabla^2 \bar B = \bar \nabla_i G^i.
\end{equation}
Here we have used
\begin{equation}
\tilde \alpha = \Psi \alpha.
\end{equation}
Eqs.~(\ref{ham3}--\ref{b3}) together with~(\ref{eul3}) form a set of
seven equations for the seven unknowns $\Psi$, $\alpha$, $G^i$, $\bar B$
and $q$.  More specifically, we have to find a solution to six coupled,
quasi-linear elliptic equation for the gravitational fields, together
with one algebraic equation for the matter. $\bar K^{ij}$ and
$\omega^i$ in the above expressions can be expressed in terms of the
unknowns with the help of Eqs.~(\ref{k2}) and~(\ref{omega}).


\subsection{The Newtonian Limit}
\label{newton}

In this Section we will briefly show that in the Newtonian limit the 
above equations approach the expected form. In particular we expect 
\begin{equation} \label{alphascale}
\alpha \rightarrow e^{\Phi} \sim 1 + \Phi,
\end{equation}
where $\Phi$ is the Newtonian potential. Also, in the Newtonian limit 
$\Phi, C, v \ll 1$, so that~(\ref{eul3}) becomes
\begin{eqnarray} 
q & = & \frac{1}{n+1} \left( C - \Phi + \frac{1}{2} v^2 \right) \nonumber \\
& = &\frac{1}{n+1} \left( C - \Phi + \frac{1}{2} \Omega^2 (x^2 + z^2) \right).
\end{eqnarray}
Here we have used $\omega^i = 0$ (absence of frame dragging in the
Newtonian limit). This limit,
by Eq.~(\ref{k2}), implies $K^{ij}=0$. With $q \ll 1$, 
Eq.~(\ref{ham3}) now reduces to
\begin{equation}
\nabla^2 \Psi = -2 \pi \Psi^5 q^n.
\end{equation}
Identifying
\begin{equation} \label{psiscale}
\Psi \rightarrow e^{-\Phi/2} \sim 1 - \frac{\Phi}{2}
\end{equation}
yields, to leading order, the Poisson equation
\begin{equation} \label{poisson}
\nabla^2 \Phi = 4 \pi \rho_0.
\end{equation}
Eq.~(\ref{lap3}) gives the same limit. 


\section{NUMERICAL METHOD}
\label{numerics}

\subsection{Constructing Quasiequilibrium Models}

Corotating, equal mass binaries in circular orbits form a
two-parameter family (just like single, uniformly rotating stars).  A
particular configuration is uniquely determined by two independent
parameters.  For computational purposes it is particularly convenient
to choose these parameters to be the maximum density $q_{\rm max}$ and
the relative separation of the stars~\cite{fn2}.

As mentioned in Section~\ref{equations}, we choose the stars to orbit
in the $y=0$ plane and to be aligned with the $z$-axis. In this case
the surface of one star will intersect the $z$-axis at two different
places. We will label the intersection closer to the origin of the
coordinate system $\bar r_A$ and the one further out $\bar r_B$. The
ratio
\begin{equation}
z_A \equiv \bar r_A/\bar r_B
\end{equation}
then parametrizes the relative separation of the stars. We can
construct an algorithm for solving the gravitational and matter
equations by modifying the algorithm used by several authors for single
rotating stars~\cite{h86,keh89,cst92}.  Making this algorithm stable
requires rescaling the coordinates with respect to $\bar r_B$ so that
\begin{equation}
\hat x = \bar x / \bar r_B~~~~~~~~~~\hat y = \bar y / \bar r_B~~~~~~~~~~
\hat z = \bar z / \bar r_B,
\end{equation}
which means that the outer edge of the matter will always be at
$\hat r_B = 1$.  We also rescale
\begin{equation}
\hat K_{ij} = \bar K_{ij} \bar r_B~~~~~~~\hat B = \bar B /\bar r_B~~~~~~~
\hat \Omega = \bar \Omega \bar r_B.
\end{equation}
Eqs.~(\ref{eul3}) and~(\ref{ham3}--\ref{b3}) are left unchanged,
except that the matter source terms in~(\ref{ham3}--\ref{g3}) have to
be multiplied by $\bar r_B^2$ and $\bar \nabla_i$ has to be replaced
by $\hat \nabla_i$. This rescaling then allows us to determine 
$\bar r_B$ as well as the the angular velocity $\hat \Omega$ and the matter
constant $C$ via an iteration process that uses $q_{max}$ and $z_A$ as the
two input parameters.

The iteration scheme starts with an initial guess for the rest density
distribution. We chose the density profile of an isolated,
spherical star, i.e. we integrate the Tolman-Oppenheimer-Volkoff
equations for the central density $q_{\rm max}$ and rescale the profile
such that it fits between $\hat r_A = z_A$ and $\hat r_B = 1$. For this
matter distribution we can then find a solution to the field
equations~(\ref{ham3}--\ref{b3}) using a full approximation storage
multigrid scheme (see, e.g.,~\cite{ptvf92}).

Once a solution to the field equations has converged to an adequate
accuracy on the finest level of the grid hierarchy, we evaluate
Eq.~(\ref{eul3}) at three different locations to find new values for
the constants $\hat \Omega$, $C$ and $\bar r_B$ as well as a new density
distribution. To do so we first search for the maximum density along
the $z$-axis~\cite{fn3} and call this location $\hat r_C$.  We can
then evaluate Eq.~(\ref{eul3}) at the three points $\hat r_A$, $\hat
r_B$ and $\hat r_C$
\begin{equation} \label{eul4}
(1+(n+1)q) \left( \alpha^2 - \Psi^4 
	(\hat \Omega \hat z - w^x)^2 \right)^{1/2} = 1 + C,
\end{equation}
where we have used $x = y = \omega^y = \omega^z = 0$ on the $z$-axis.
Note that at $\hat r_A$ and $\hat r_B$ the density vanishes $q = 0$.
This set, on first sight, looks like three equations for the two unknows
$\Omega$ and $C$. However, changing the scaling parameter $\bar r_B$ will
also change the gravitational fields, so that $\alpha$
and $\Psi$ will implicitly depend on $\bar r_B$. We determine
how $\alpha$ and $\Psi$ scale from the Newtonian
limit. Rescaling the Poisson equation shows that the Newtonian
potential $\Phi$ scales with $\bar r_B^2$. Eqs.~(\ref{alphascale})
and~(\ref{psiscale}) therefore suggest that $\alpha$ and $\Psi$ should
be rescaled according to
\begin{equation}
\alpha = (\hat \alpha)^{\bar r_B^2}~~~~~~~~~~
\Psi = (\hat \Psi)^{-\bar r_B^2/2}.
\end{equation}
Inserting these scale relations into~(\ref{eul4}) then yields three
equations for the three constants $\hat \Omega$, $C$ and $\bar r_B$, which can
be solved iteratively.  Once the constants have been determined the
new matter distribution can be calculated using Eq.~(\ref{eul3}).

The iteration can then be continued by finding the new fields for the
new matter distribution. At each step we calculate the residuals of
Eqs.~(\ref{ham3}--\ref{b3}) and integrate these over the numerical
grid. We typically stop the iteration when the sum of these six
integrated residuals is smaller than about 1 \% of the
estimated truncation error on the finest grid.

Once an iteration has been completed, we can calculate several physical
quantities that characterize the configuration. The total rest-mass
$M_{0,\rm tot}$ is
\begin{equation}
M_{0,\rm tot} = \int_{\cal M} \rho_0 u^{\alpha} d^3 \Sigma_{\alpha}
	= \int_{\cal M} \rho_0 u^t \sqrt{-g} d^3 x,
\end{equation}
where the subscript ${\cal M}$ denotes integration over the support of the
matter and $\sqrt{-g} = \alpha \Psi^{6}$.  In nondimensional form we
can therefore write
\begin{equation} \label{M_0}
\bar M_{0,\rm tot} \equiv \kappa^{-n/2} M_{0,\rm tot} 
	= \bar r_B^3 \int_{\cal M} \alpha \Psi^6 u^t q^n d^3 \hat x.
\end{equation}
The total mass-energy (ADM mass) is
\begin{equation}
M_{\rm tot} = - \frac{1}{2 \pi} \oint_{\infty} \nabla^i \Psi d^2 S_i
	=  - \frac{1}{2 \pi} \int_{\infty} \nabla^2 \Psi d^3 x,
\end{equation}
Using the Hamiltonian constraint~(\ref{ham3}) this can be rewritten
\begin{eqnarray}
M_{\rm tot} & = & \frac{1}{16 \pi} \int_{\infty} \Psi^{-7} 
	\tilde K_{ij} \tilde K^{ij} d^3 x
	\nonumber \\
	& &+ \int_{\cal M} \Psi^5 q^n \left( \frac{1+(1+n)q}{1-v^2} - q \right) d^3 x,
\end{eqnarray}
or, in nondimensional form,
\begin{eqnarray} \label{M}
\bar M_{\rm tot} & \equiv & \kappa^{-n/2} M_{\rm tot} =
	\frac{\bar r_B}{16 \pi} 
	\int_{\infty} \Psi^{-7} \hat K_{ij} \hat K^{ij} d^3 \hat x
	\nonumber \\
	& & + \bar r_B^3 \int_{\cal M} 
	\Psi^5 q^n \left( \frac{1+(1+n)q}{1-v^2} - 
	q \right) d^3 \hat x.
\end{eqnarray}
Eq.~(\ref{M}) is the actual form we use to evaluate $M_{\rm tot}$.
The angular momentum is aligned with the $y$-axis and 
can be defined as
\begin{equation}
J_{\rm tot} = \frac{\epsilon_{yjk}}{8 \pi} \oint_{\infty} 
	x^j \tilde K^{kl} d^2 S_i
 	= \frac{\epsilon_{yjk}}{8 \pi} \int_{\infty} x^j 
	\nabla_l \tilde K^{kl} d^3 x
\end{equation}
(see, e.g.,~\cite{by80}). This is the total angular momentum contained
in the spacetime and includes both the orbital and spin angular
momentum of the stars.  Using $\nabla_l \tilde K^{kl} = \Psi^{10}
D_l K^{kl}$ as well as the momentum constraint~(\ref{mom1}), this
can be rewritten
\begin{equation}
J_{\rm tot} = \int_{\cal M} \Psi^{10} (z j^x - x j^z)\, d^3 x = 
	 \int_{\cal M} \Psi^{10} f_{ij} \xi^i j^j d^3 x,
\end{equation}
where we have also used definition~(\ref{xi}). Finally, we can
substitute~(\ref{jsource}) for $j^i$ and write the angular momentum in
the nondimensional form
\begin{eqnarray} \label{J}
\bar J_{\rm tot} & &\equiv \kappa^{-n} J_{\rm tot} = \nonumber \\
& &	\bar r_B^4 \int_{\cal M} \frac{\Psi^{10}}{\alpha} q^n 
	\frac{1 + (1+n)q}{1 - v^2} 
	f_{ij} \hat \xi^{i} (\hat \Omega \hat \xi^j - \omega^j) \,d^3 \hat x,
\end{eqnarray}
where we have rescaled $\bar \xi$ according to 
$\hat \xi = \bar \xi /\bar r_B$.

In the following we will denote half the total rest-mass, mass and
angular momentum by $\bar M_0 = \bar M_{0,\rm tot}/2$, $\bar M = \bar
M_{\rm tot}/2$ and $\bar J = \bar J_{\rm tot}/2$.  In the limit of
large separation, $\bar M_0$ and $\bar M$ approach the corresponding 
values of isolated stars.

Performing numerical simulations in three dimensions requires large
computational resources. We have therefore implemented our algorithm in
a parallel environment using the DAGH infrastructure~\cite{pb95} and
run it both on the SP2 cluster at the Cornell Theory Center and the
Origin2000 at the National Center for Supercomputing Applications at
the University of Illinois.  We typically use grids of $(64)^3$ or
$(128)^3$ gridpoints, and run the code in parallel on 8 processors.
DAGH has been developed as part of the Binary Black Hole Grand
Challenge Project and is a package of routines and computational
structures that allows for a convenient implementation of parallel
applications on grid hierarchies.


\subsection{Constructing Quasiequilibrium Sequences}
\label{numerics-seq}

In~addition~to~constructing~individual~quasi\-equilibrium
configurations, we can also build quasiequilibrium sequences of
constant rest-mass $\bar M_0$. As we will discuss in
Section~\ref{results}, these sequences provide approximate
evolutionary tracks of inspiraling neutron star binaries.

Our quasiequilibrium configurations are parametrized by their relative
separation $z_A$ and maximum density $q_{\rm max}$. We therefore have
to find a path through this two-dimensional parameter space along
which $\bar M_0$ is constant. This can be achieved in several
different ways. For example, for each separation $z_A$ one could vary
$q_{\rm max}$ until a configuration of mass $\bar M_0$ has been
found~\cite{cst92}. Here we found it easier to start with a small (and
hence only mildly relativistic) $q_{\rm max}$ for each $z_A$, and then
increment $q_{\rm max}$ in small steps keeping $z_A$ constant. The
results can be tabulated, and the procedure repeated for a different
$z_A$. Once sufficient data have been collected one can then
interpolate to a chosen rest-mass $\bar M_0$. Note that for each $z_A
= {\rm const}$ sequence we adjusted the outer boundary so that the
number of gridzones covering the stars is the same for all
separations.

We have performed several tests to check our code.  In two different
regimes the results can be compared with known solutions: for small
masses and weak fields we recover the Newtonian limit, and for large
separations we approach the Oppenheimer-Volkoff spherical solution for
each star and its near-by field. We have also checked the fully
relativistic identity~\cite{h70,b72}
\begin{equation} \label{first_law}
dM_{\rm tot} = \Omega dJ_{\rm tot},
\end{equation}
which holds along constant rest-mass sequences. To evaluate
Eq.~(\ref{first_law}), we have to take numerical differences between
integrals of very similar magnitude, so that their relative error was
much larger than that of the individual integrals. Nevertheless, we
found that this identity is satisfied typically to $\sim 10$\% (except
close to turning points, where the error due to the differentiation
dominates). We expect that the numerical data presented in this paper
are typically accurate to within a few percent, and are confident that
our code correctly predicts qualitative features, like, for example,
changes in the maximum allowed mass.


\section{RESULTS}
\label{results}

\subsection{Sequences for $n = 1.0$}
\label{n=1.0}

\begin{figure}
\epsfxsize=3in
\begin{center}
\leavevmode
\epsffile{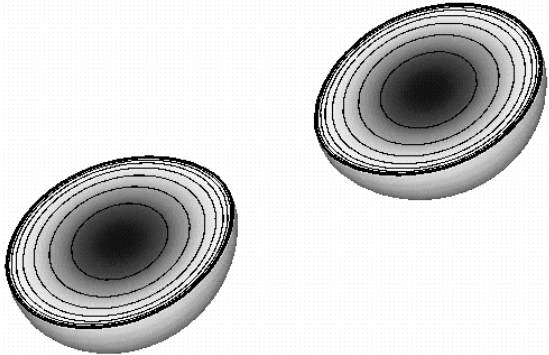}
\end{center}
\caption{Rest-density contours in the equatorial plane for a neutron 
star binary close to the ISCO. Each star has a rest-mass of $\bar M_0 =
0.169$, corresponding to a compaction in isolation of $(M/R)_{\infty}
= 0.175$. The contours show isosurfaces of the rest-density in
decreasing factors of 0.556.}
\end{figure}

In this Section we discuss configurations and sequences with a
polytropic index $n = 1$, representing a fairly stiff equation of
state. This is a particularly interesting example, since realistic
neutron stars are expected to be governed by equations of state of
similar stiffness.  Results for $n=1.5$ and $n=2$ will be presented in
Section~\ref{otherns}. Numerical values in geometrized units can be
obtained from our nondimensional ``barred'' quantities by multiplying
with appropriate powers of $\kappa$, according to Eqs.~(\ref{M_0}),
(\ref{M}) and (\ref{J}) (for example $M = \kappa^{n/2}
\bar M$, $J = \kappa^n \bar J$, and $\rho_0 = \kappa^{-n} \bar \rho_0$).

In Fig.~1 we show the density profile in the equatorial plane of a
binary neutron star. Here $z_A = 0.175$, and the stars are close to
the ISCO (see below). Each star has a rest-mass of $\bar M_0 = 0.169$,
corresponding to a compaction in isolation of $(M/R)_{\infty} =
0.175$. The contours show isosurfaces of the rest-density in
decreasing factors of 0.556.  The maximum compaction of a stable
$n=1.0$ polytrope in isolation is $(M/R)_{\infty} = 0.216$,
corresponding to a maximum rest-mass $\bar M_0 = 0.180$ and a maximum
mass $\bar M = 0.164$.

\begin{figure}
\epsfxsize=3in
\begin{center}
\leavevmode
\epsffile{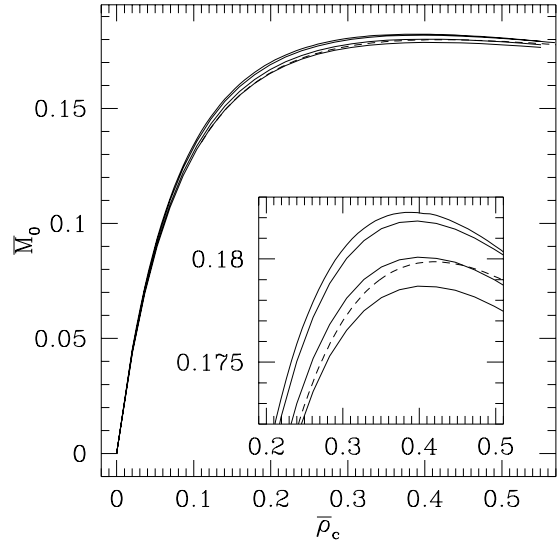}
\end{center}
\caption{Rest-Mass $\bar M_0$ versus maximum density $\bar \rho_{c}$ for
separations $z_A = 0.3$ (bottom solid line), 0.2, 0.1 and 0.0 (top
line).  The dashed line is the Oppenheimer-Volkoff result for a
$n=1.0$ polytrope. The insert is a blow-up of the region around the
maximum mass.}
\end{figure}

In Fig.~2 we plot the rest-mass $\bar M_0$ versus the maximum density
$\bar \rho_c = \bar \rho_0^{\rm max} + \bar \rho_i^{\rm max}$
for several different separations between $z_A = 0.3$ (roughly two
stellar radii apart) and $z_A = 0$ (touching). As $z_A \rightarrow 1$,
we expect these curves to approach the spherical Oppenheimer-Volkoff
(OV) result, which we included as the dashed line in Fig.~2. Note,
however, that the exact OV curve is computed from a one-dimensional
ordinary differential equation with very high accuracy, while the
binary configurations have been calculated on very coarse,
three-dimensional numerical grids. From convergence tests we know that
we systematically underestimate masses, and accordingly, for large
separations, we find masses slightly smaller than the corresponding OV
masses.  All graphs lie within less than 2\% of the OV curve, showing
that the presence of a companion star has only very little influence
on the mass-density relationship.

As we decrease the separation, the mass supported by a given central
density $\bar \rho_c$ increases slightly. In particular, the maximum
rest-mass increases from $\bar M_0^{\rm max} = 0.179$ for $z_A = 0.3$
to $\bar M_0^{\rm max} = 0.182$ for stars in contact. This trend
clearly suggests that {\em the maximum allowed mass of neutron stars
in close binaries is slightly larger than in isolation}. This increase
is caused in part by the rotation of the stars and in part by the
tidal fields. More specifically, we find that the increase of the
maximum allowed mass is comparable to the corresponding increase of an
isolated neutron star rotating with the same angular
velocity~\cite{cst92}. Any destabilizing, relativistic effect in
binaries therefore has to be smaller.

\begin{figure}
\epsfxsize=3in
\begin{center}
\leavevmode
\epsffile{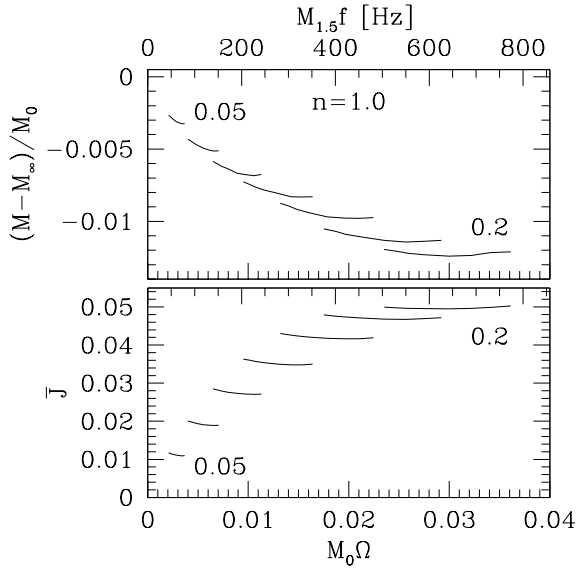}
\end{center}
\caption{Binding energy and angular momentum as a function of the
angular velocity for several different values of $\bar M_0$. The 
curves are labeled by the compaction $(M/R)_{\infty}$ of the stars
in isolation at infinity, starting with 0.05 and increasing in 
steps of 0.0025 up to 0.2. The maximum compaction of a stable, 
isolated, non-rotating $n=1.0$ polytrope is 0.217. The upper label
gives the orbital frequency for stars with a rest-mass of 
$1.5 M_{\odot}$}
\end{figure}

The collapse of binary neutron stars to black holes prior to merger
reported by WMM could, in principle, be caused either by a decrease of
the maximum allowed mass, or by a dynamical instability. As we have
shown, the maximum allowed mass, within our assumptions and
approximations, increases, which rules out the first possibility.
Note, however, that we are only constructing quasiequilibrium
configurations, which may not be dynamically stable.
In~\cite{bcsst97b} we show that all inspiraling binary neutron stars
are {\em secularly} stable against radial collapse to black holes all
the way down to the innermost stable circular orbit (ISCO). While this
does not completely rule out the existence of a {\em dynamical}
instability, we note that in Newtonian binaries, dynamical
instabilities always occur later along equilibrium sequences than
secular instabilities~\cite{c70,lrs93}. The same result has been shown
for single, rotating relativistic stars~\cite{fis88}. Recently,
Thorne~\cite{t97} has argued analytically that tidal fields stabilize
systems and that stars which are stable in isolation are stable with
respect to both secular and dynamical modes in binary configurations.

Fig.~2 demonstrates that at fixed rest-mass, the central density
decreases as the stars approach each other and get tidally
deformed. This effect, as well as the increase of the maximum allowed
mass, is consistent with post-Newtonian
predictions~\cite{l96,lrs97,w97}.

\begin{figure}
\epsfxsize=3in
\begin{center}
\leavevmode
\epsffile{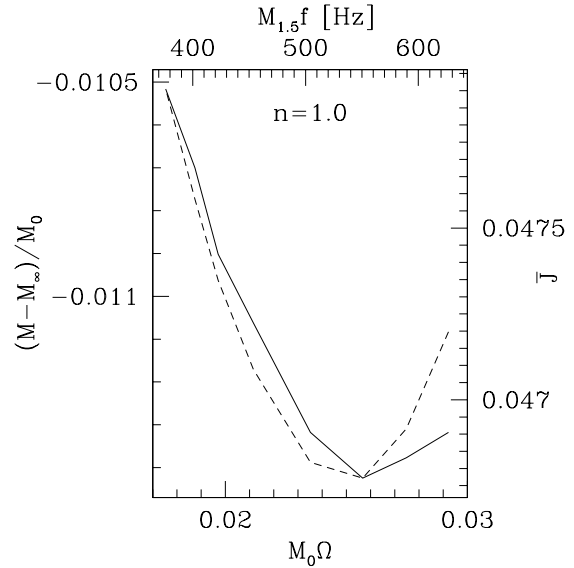}
\end{center}
\caption{Blow-up of two curves in Fig.~2:
binding energy (solid) and angular momentum (dashed) as a function of the
angular velocity for a binary with $\bar M_0 = 0.169$ and
$(M/R)_{\infty} = 0.175$.}
\end{figure}

Next we construct sequences of constant rest-mass $\bar M_0$, which up
to the ISCO approximate evolutionary sequences. As discussed in
Section~\ref{assumptions}, we maintain corotation, whereas in reality
it is more likely that circulation will be conserved.  Nevertheless,
our sequences are the first sequences of inspiraling binaries in full
general relativity. Moreover, post-Newtonian sequences of constant
circulation are not vastly different from corotating
sequences~\cite{lrs97}.  In Fig.~3 we plot the binding energy $(M -
M_{\infty})/M_0$ and the angular momentum $\bar J$ as a function of
separation for several different rest-masses.  Since the separation is
not an invariant quantity, we have parametrized the sequence by the
nondimensionalized angular velocity $M_0 \Omega$ ($= \bar M_0 \bar
\Omega$).  Our curves do not connect to $M_0 \Omega = 0$,
corresponding to infinite separation, since we can numerically resolve
only fairly close models.

In the top half of Fig.~3 we show plots for sequences for several
different, increasingly relativistic rest-masses. The curves are
labeled by the compaction $(M/R)_{\infty}$ that the stars would have
in isolation at infinity. We have plotted graphs for $(M/R)_{\infty}$
between 0.05 and 0.2 in increments of 0.025.  In the lower half of
Fig.~3 we show corresponding plots of $\bar J$.  According to
Eq.~(\ref{first_law}) the minima in both curves must agree, which
they do within our numerical accuracy. In Fig.~4 we show a blow-up of
the two curves for stars with $(M/R)_{\infty} = 0.175$.

For infinitely separated stars, both the binding energy and the
angular velocity vanish. As the stars approach each other, the angular
velocity increases while the binding energy decreases. This effect is
essentially Newtonian and is even evident for two Newtonian point
masses.  As the stars approach, however, finite size effects
eventually play an important role. The energy associated with the
rotation of the individual stars adds to the (negative) binding
energy, and therefore reduces it.  For stiff enough equations of
state, for which the moment of inertia and hence the rotational energy
of the individual stars is large (see Section~\ref{otherns}),
the binding energy goes through a minimum and then increases again
prior to contact.  The location of the minimum marks the onset of a
secular instability, beyond which the binary can no longer maintain
corotation. It is expected that the dynamical instability
defining the ISCO occurs after, but close to the onset of the secular
instability~\cite{c70,lrs93}. In the following we will refer
to the location of the minimum as the ISCO.

The upper labels give the orbital frequency in Hz for stars of
rest-mass $1.5 M_{\odot}$. The corresponding gravitational wave
frequency is larger by a factor of 2 for the dominant quadrupole mode.
For small values of the compaction we find ISCO frequencies comparable
to those reported by WMM. However, for larger compaction and more
relativistic configurations we find frequencies very similar to what
is found from post-Newtonian calculations~\cite{kww93}.

\begin{table}
\begin{center}
\begin{tabular}{ccccc}
$\bar M_0$ & $\bar M_{\infty}$ & $(M/R)_{\infty}$ & $M_0
	\Omega_{ISCO}$ & $(J_{\rm tot}/M_{\rm tot}^2)_{ISCO}$ \\
\tableline
0.059	& 0.058	& 0.05	& 0.003 & 1.69 \\
0.087	& 0.084 & 0.075 & 0.0065 & 1.37 \\ 
0.112	& 0.106	& 0.1	& 0.01	& 1.22 \\
0.134	& 0.126	& 0.125	& 0.015	& 1.12 \\
0.153	& 0.142	& 0.15	& 0.02	& 1.05 \\
0.169	& 0.155	& 0.175	& 0.025	& 1.00 \\
0.178 	& 0.162 & 0.2	& 0.03	& 0.97 
\end{tabular}
\end{center}
\caption{Numerical values for sequences of constant rest-mass $\bar M_0$
and polytropic index $n=1$.  We tabulate the total energy $\bar
M_{\infty}$ and compaction $(M/R)_{\infty}$ each star would have in
isolation as well as the angular velocity $M_0 \Omega$ and the angular
momentum $J_{\rm tot}/M_{\rm tot}^2$ at the ISCO. The maximum rest-mass 
in isolation is $\bar M_0^{\rm max} = 0.180$.}
\end{table}

We summarize our results in Table 2, where we also include the
dimensionless angular momentum $J_{\rm tot}/M_{\rm tot}^2 = J/2M^2$
at the ISCO. For small rest-masses, this value is larger than unity,
in agreement with WMM. For high enough rest-masses, however, it drops
below unity, so that the two stars could plunge and form a Kerr black
hole without having to lose additional angular momentum.


\subsection{Sequences for $n = 1.5$ and $n = 2.0$}
\label{otherns}

\begin{figure}
\epsfxsize=3in
\begin{center}
\leavevmode
\epsffile{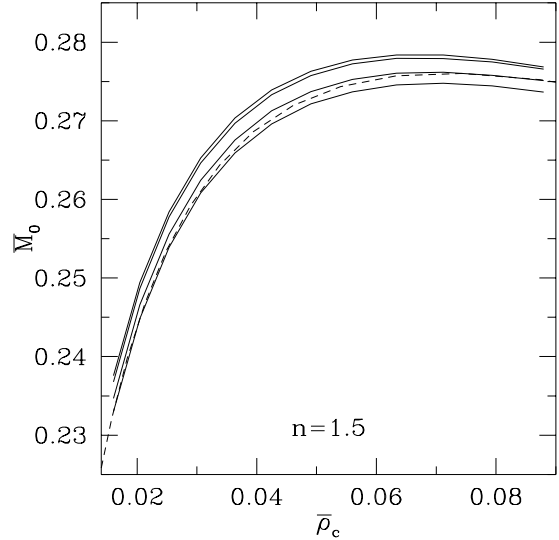}
\end{center}
\caption{Rest-Mass $\bar M_0$ of a $n=1.5$ polytrope versus maximum
density $\bar \rho_c$ for separations $z_A = 0.3$ (bottom solid line),
0.2, 0.1 and 0.0 (top line). The dashed line is the
Oppenheimer-Volkoff result.}
\end{figure}

In this Section we will present results for polytropic indices of $n =
1.5$ and 2.0, representing softer equations of state. Except for the
absence of an ISCO prior to contact (see below) all results are
qualitatively very similar to those for $n=1$. In particular, we
consistently find a decrease of the maximum density as the stars
approach, and an increase in the maximum allowed mass. The relative
size of these effects differs for three basic reasons: First, for
softer equations of state, the maximum mass of a star occurs at a
smaller value of the compaction $M/R$, and hence relativistic effects
play a smaller role. Second, for softer equations of state these stars
are more centrally condensed.  We therefore expect tidal fields to
play a less important role for the stability of these stars in close
binaries. While it is easier to deform their surface, the bulk of the
matter is very concentrated at the core of the stars and well shielded
from the tidal field of the companion. Third, for softer equations of
state, the stars have a smaller orbital frequency even at very small
separations, so that the effects of rotation are smaller.
Accordingly we find that the maximum allowed mass still increases with
decreasing separation, but the effect is smaller than for $n=1$.

\begin{figure}
\epsfxsize=3in
\begin{center}
\leavevmode
\epsffile{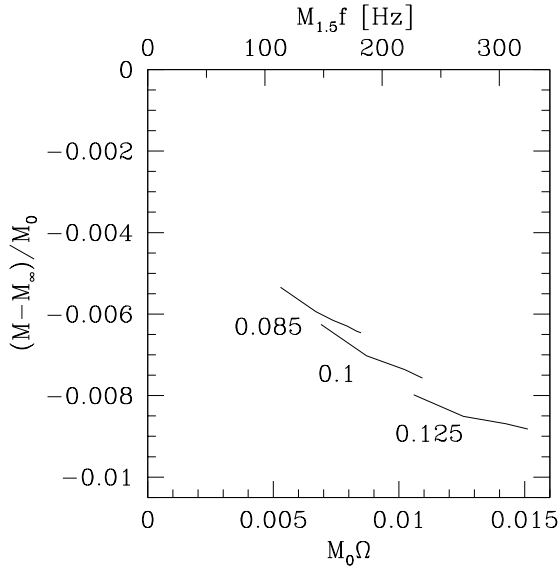}
\end{center}
\caption{Binding energy of $n=1.5$ polytropes 
as a function of the angular velocity for different rest-masses. The 
curves are labeled by the compaction $(M/R)_{\infty}$ of the stars in
isolation. The maximum compaction for a stable, isolated, nonrotating
$n=1.5$ polytrope is 0.136.}
\end{figure}

More centrally condensed stars have a smaller moment of inertia, and
hence the rotational kinetic energy associated with the spin of the
stars is smaller than for less centrally condensed stars. Therefore a
turning point in the binding energy curve can only be expected for
stars with a stiff enough equation of state. This effect has been
discussed by several authors in the context of Newtonian
theory~\cite{h86,lrs93,rs94}.  We did not see a turning point for $n
\geq 1.5$, in agreement with~\cite{h86,rs94}.  For these polytropic
indices there is no ISCO prior to contact, and we expect the orbits to
be stable until the stars touch and form a contact binary. This is the
first construction of a contact binary in full general
relativity. Proving the existence of a contact binary neutron star
(by, e.g., the signature of its gravitational waveform) would indicate
that the equation of state of nuclear matter is rather soft. We do not
expect this to be the case~\cite{st83}.

In Fig.~5 we plot the rest-mass versus the central density for several
different separations for $n=1.5$. Qualitatively the result is very
similar to Fig.~2 for $n=1$: For all separations the curve differs
from the OV result by less than 1\%. For decreasing separation we find
a small increase in the allowed mass that a given density can support.
In particular, the maximum quasiequilibrium rest-mass increases by roughly
1.2\% from $\bar M = 0.275$ for $z_A = 0.3$ to 0.278 for stars in
contact.  For $n = 1$ the corresponding increase is about 2 \%.  The
maximum density decreases as the stars approach and get tidally
deformed.

\begin{figure}
\epsfxsize=3in
\begin{center}
\leavevmode
\epsffile{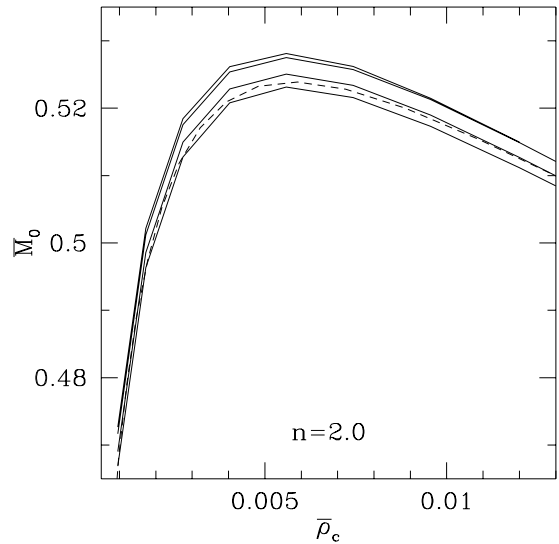}
\end{center}
\caption{Same as Fig.~5 for a $n=2.0$ polytrope.}
\end{figure}

In Fig.~6 we plot the binding energy of $n=1.5$ polytropes as a
function of the angular velocity. We show results for several
different rest-masses and label them by the compaction
$(M/R)_{\infty}$ for the same stars in isolation. In
contrast to the results for $n=1$, these curves no longer show a
turning point. This implies that the stars are secularly stable all
the way to touching.

\begin{figure}
\epsfxsize=3in
\begin{center}
\leavevmode
\epsffile{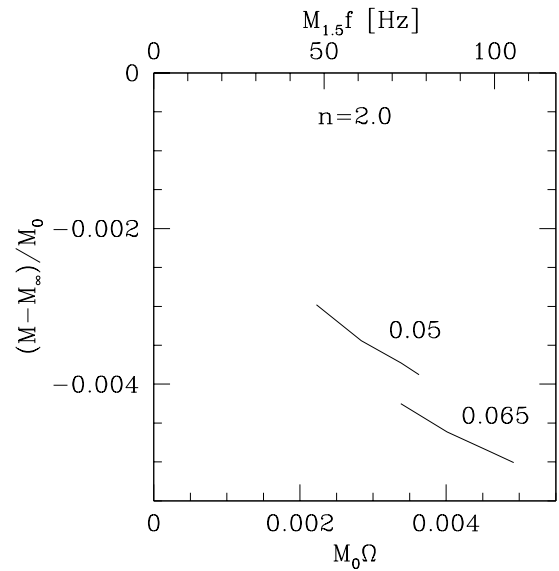}
\end{center}
\caption{Same as Fig.~6 for a $n=2.0$ polytrope. The maximum compaction
$(M/R)_{\infty}$ for a stable, isolated, nonrotating
$n=1.5$ polytrope is 0.075.}
\end{figure}

In Figs.~7 and 8 we show the corresponding results for $n=2$
polytropes.  Again, in Fig.~7 we show the rest-mass versus central
density. The maximum quasiequilibrium rest-mass increases from $\bar M =
0.523$ for $z_A = 0.3$ to 0.528 for touching stars.  This relative
increase of roughly 1 \% is smaller than even for $n=1.5$. As expected,
the binding energies in Fig.~8 do not show a turning point, so that
the binaries are secularly stable all the way to touching.


\section{SUMMARY AND CONCLUSIONS}
\label{summary}

We report on the first fully relativistic calculation of binary
neutron stars in quasiequilibrium. We previously presented some of
our preliminary results in~\cite{bcsst97a}; here we describe in detail
all our assumptions and approximations, equations and numerical
algorithm, as well as results for different polytropic indices.  We
integrate a subset of Einstein's equations, coupled to the equation of
hydrostatic equilibrium, to solve the initial value problem for
binaries.  We construct models of corotating binary neutron stars in
close circular orbit, including relativistic models of contact
binaries. We also construct sequences of constant rest-mass
configurations parametrized by their separation and orbital angular
frequency.

We find that the maximum density of the stars decreases as the stars
approach and get tidally deformed. Simultaneously, the mass that a
given maximum density can support increases as the stars approach. In
particular, we find that the maximum allowed mass of neutron stars in
quasiequilibrium binaries increases with decreasing separation. These
effects are larger for smaller polytropic index (and hence stiffer
equation of state).

Searching for turning points of the binding energy of constant rest
mass sequences, we locate, for small enough polytropic index, the
ISCO. As in the case of Newtonian configurations, an ISCO exists only for
indices $n \stackrel{<}{\sim} 1.5$; for softer equations of state,
contact is reached prior to the onset of orbital instability.

In~\cite{bcsst97b} we presented a more careful analysis of the radial
stability of relativistic binary neutron stars against collapse. We
showed that all inspiraling binary neutron stars are secularly stable
against radial collapse to black holes all the way to the ISCO (or
contact, if, for large enough $n$, no ISCO is encountered). We do not
find any evidence for a destabilization of neutron stars in close
binary orbits.

\acknowledgments

It is a pleasure to thank Manish Parashar for his help with the
implementation of DAGH, and Andrew Abrahams, James Lombardi and Fred
Rasio for several helpful discussions. We would also like to thank
Matthew Duez, Eric Engelhard and John Fregeau for helping with the
visualization of our data and the production of Fig.~1.  This work was
supported by NSF Grant AST 96-18524 and NASA Grant NAG 5-3420 at
Illinois, NSF Grant PHY 94-08378 at Cornell, and by the NSF Binary
Black Hole Grand Challenge Grant Nos.~NSF PHY 93-18152/ASC 93-18152
(ARPA supplemented). Computations were performed at the Cornell Center
for Theory and Simulation in Science and Engineering and the National
Center for Supercomputing Applications, University of Illinois at
Urbana-Champaign.

\begin{appendix}
\section{Numerical Results for Selected Sequences}

In the following we tabulate numerical values for selected sequences.
For a given polytropic index $n$ and the rest-mass (baryon mass) $\bar
M_0$ of one star (or equivalently its compaction in isolation
$(M/R)_{\infty}$), we list the relative separation 
$z_A = \bar r_A/\bar r_B$,
the maximal density parameter $q^{\rm max}$, the mass $\bar M$
the angular momentum $\bar J$, the (orbital)
frequency $\bar \Omega$, and the locations $\bar r_A$, $\bar r_B$ and
$\bar r_C$. We have ``barred'' these quantities as a reminder that
they are dimensionless coordinate values. Recall that $\rho$, $\rho_0$
and $P$ may be obtained from $q$ via Eqs.~(\ref{rho_0}) --~(\ref{P}).

\begin{table}[h]
\begin{center}
$n = 1$, $\bar M_0 = 0.0595$, $(M/R)_{\infty} = 0.05$
\begin{tabular}{cccccccc}
$z_A$ & $q^{\rm max}$ & $\bar M$ & $\bar J$ & $\bar \Omega$ &
    $\bar r_A$ & $\bar r_B$ & $\bar r_C$ \\
\tableline
  0.00 & 0.0275  & 0.057806 & 0.01095 & 0.061 & 0.000 & 1.529 & 2.773\\
  0.10 & 0.0278  & 0.057806 & 0.01094 & 0.057 & 0.281 & 1.594 & 2.810\\
  0.15 & 0.0281  & 0.057809 & 0.01098 & 0.053 & 0.430 & 1.677 & 2.868\\
  0.20 & 0.0284  & 0.057815 & 0.01109 & 0.048 & 0.591 & 1.791 & 2.959\\
  0.25 & 0.0286  & 0.057825 & 0.01129 & 0.042 & 0.771 & 1.940 & 3.087\\
  0.30 & 0.0288  & 0.057836 & 0.01155 & 0.037 & 0.975 & 2.118 & 3.251\\
\end{tabular}
\end{center}
\caption{}
\end{table}

\begin{table}[h]
\begin{center}
$n = 1$, $\bar M_0 = 0.1118$, $(M/R)_{\infty} = 0.1$
\begin{tabular}{cccccccc}
$z_A$ & $q^{\rm max}$ & $\bar M$ & $\bar J$ & $\bar \Omega$ &
    $\bar r_A$ & $\bar r_B$ & $\bar r_C$ \\
\tableline
  0.00 & 0.0658  & 0.105511 & 0.02715 & 0.101 & 0.000 & 1.289 & 2.353\\
  0.10 & 0.0667  & 0.105502 & 0.02707 & 0.094 & 0.238 & 1.346 & 2.384\\
  0.15 & 0.0676  & 0.105509 & 0.02710 & 0.087 & 0.365 & 1.418 & 2.433\\
  0.20 & 0.0685  & 0.105521 & 0.02729 & 0.079 & 0.502 & 1.516 & 2.511\\
  0.25 & 0.0693  & 0.105558 & 0.02766 & 0.070 & 0.655 & 1.644 & 2.621\\
  0.30 & 0.0698  & 0.105593 & 0.02818 & 0.062 & 0.828 & 1.797 & 2.763\\
\end{tabular}
\end{center}
\caption{}
\end{table}

\begin{table}[h]
\begin{center}
$n = 1$, $\bar M_0 = 0.1341$, $(M/R)_{\infty} = 0.125$
\begin{tabular}{cccccccc}
$z_A$ & $q^{\rm max}$ & $\bar M$ & $\bar J$ & $\bar \Omega$ &
    $\bar r_A$ & $\bar r_B$ & $\bar r_C$ \\
\tableline
  0.00 & 0.0912  & 0.124786 & 0.03496 & 0.122 & 0.000 & 1.172 & 2.148\\
  0.10 & 0.0926  & 0.124785 & 0.03482 & 0.114 & 0.217 & 1.225 & 2.175\\
  0.15 & 0.0940  & 0.124787 & 0.03482 & 0.106 & 0.332 & 1.291 & 2.219\\
  0.20 & 0.0954  & 0.124819 & 0.03500 & 0.096 & 0.458 & 1.381 & 2.290\\
  0.25 & 0.0967  & 0.124849 & 0.03538 & 0.086 & 0.597 & 1.498 & 2.390\\
  0.30 & 0.0976  & 0.124898 & 0.03596 & 0.076 & 0.756 & 1.639 & 2.520\\
\end{tabular}
\end{center}
\caption{}
\end{table}

\begin{table}[h]
\begin{center}
$n = 1$, $\bar M_0 = 0.1534$, $(M/R)_{\infty} = 0.15$
\begin{tabular}{cccccccc}
$z_A$ & $q^{\rm max}$ & $\bar M$ & $\bar J$ & $\bar \Omega$ &
    $\bar r_A$ & $\bar r_B$ & $\bar r_C$ \\
\tableline
  0.00 & 0.1235  & 0.140851 & 0.04188 & 0.146 & 0.000 & 1.056 & 1.943\\
  0.10 & 0.1256  & 0.140842 & 0.04167 & 0.137 & 0.196 & 1.104 & 1.967\\
  0.15 & 0.1280  & 0.140846 & 0.04162 & 0.127 & 0.300 & 1.163 & 2.005\\
  0.20 & 0.1303  & 0.140859 & 0.04174 & 0.116 & 0.413 & 1.244 & 2.067\\
  0.25 & 0.1325  & 0.140903 & 0.04210 & 0.104 & 0.539 & 1.350 & 2.156\\
  0.30 & 0.1341  & 0.140971 & 0.04268 & 0.092 & 0.682 & 1.477 & 2.273\\
\end{tabular}
\end{center}
\caption{}
\end{table}

\begin{table}[h]
\begin{center}
$n = 1$, $\bar M_0 = 0.1685$, $(M/R)_{\infty} = 0.175$
\begin{tabular}{cccccccc}
$z_A$ & $q^{\rm max}$ & $\bar M$ & $\bar J$ & $\bar \Omega$ &
    $\bar r_A$ & $\bar r_B$ & $\bar r_C$ \\
\tableline
  0.00 & 0.1647  & 0.152893 & 0.04719 & 0.173 & 0.000 & 0.944 & 1.743\\
  0.10 & 0.1683  & 0.152883 & 0.04691 & 0.163 & 0.176 & 0.987 & 1.762\\
  0.15 & 0.1726  & 0.152875 & 0.04677 & 0.152 & 0.268 & 1.038 & 1.792\\
  0.20 & 0.1769  & 0.152893 & 0.04681 & 0.139 & 0.368 & 1.108 & 1.844\\
  0.25 & 0.1811  & 0.152936 & 0.04708 & 0.125 & 0.480 & 1.201 & 1.920\\
  0.30 & 0.1844  & 0.152997 & 0.04758 & 0.111 & 0.606 & 1.312 & 2.022\\
\end{tabular}
\end{center}
\caption{}
\end{table}

\begin{table}[h]
\begin{center}
$n = 1$, $\bar M_0 = 0.1781$, $(M/R)_{\infty} = 0.2$
\begin{tabular}{cccccccc}
$z_A$ & $q^{\rm max}$ & $\bar M$ & $\bar J$ & $\bar \Omega$ &
    $\bar r_A$ & $\bar r_B$ & $\bar r_C$ \\
\tableline
  0.00 & 0.2164  & 0.160183 & 0.05024 & 0.202 & 0.000 & 0.841 & 1.560\\
  0.10 & 0.2228  & 0.160174 & 0.04989 & 0.191 & 0.157 & 0.877 & 1.572\\
  0.15 & 0.2327  & 0.160137 & 0.04963 & 0.180 & 0.238 & 0.917 & 1.587\\
  0.20 & 0.2450  & 0.160130 & 0.04948 & 0.168 & 0.323 & 0.970 & 1.616\\
  0.25 & 0.2590  & 0.160145 & 0.04953 & 0.154 & 0.415 & 1.038 & 1.662\\
  0.30 & 0.2741  & 0.160189 & 0.04975 & 0.139 & 0.517 & 1.119 & 1.725\\
\end{tabular}
\end{center}
\caption{}
\end{table}

\begin{table}[h]
\begin{center}
$n = 1.5$, $\bar M_0 = 0.241$, $(M/R)_{\infty} = 0.85$
\begin{tabular}{cccccccc}
$z_A$ & $q^{\rm max}$ & $\bar M$ & $\bar J$ & $\bar \Omega$ &
    $\bar r_A$ & $\bar r_B$ & $\bar r_C$ \\
\tableline
  0.00 & 0.0626  & 0.231583 & 0.13408 & 0.035 & 0.000 & 3.409 & 6.227\\
  0.10 & 0.0633  & 0.231623 & 0.13471 & 0.032 & 0.631 & 3.569 & 6.318\\
  0.20 & 0.0650  & 0.231708 & 0.13738 & 0.027 & 1.328 & 4.014 & 6.642\\
  0.30 & 0.0665  & 0.231853 & 0.14341 & 0.021 & 2.184 & 4.740 & 7.281\\
\end{tabular}
\end{center}
\caption{}
\end{table}

\begin{table}
\begin{center}
$n = 1.5$, $\bar M_0 = 0.258$, $(M/R)_{\infty} = 0.1$
\begin{tabular}{cccccccc}
$z_A$ & $q^{\rm max}$ & $\bar M$ & $\bar J$ & $\bar \Omega$ &
    $\bar r_A$ & $\bar r_B$ & $\bar r_C$ \\
\tableline
  0.00 & 0.0794  & 0.246547 & 0.14275 & 0.042 & 0.000 & 3.038 & 5.561\\
  0.10 & 0.0802  & 0.246600 & 0.14346 & 0.039 & 0.564 & 3.183 & 5.643\\
  0.20 & 0.0831  & 0.246688 & 0.14574 & 0.033 & 1.180 & 3.564 & 5.902\\
  0.30 & 0.0855  & 0.246887 & 0.15166 & 0.026 & 1.938 & 4.204 & 6.460\\
\end{tabular}
\end{center}
\caption{}
\end{table}

\begin{table}
\begin{center}
$n = 1.5$, $\bar M_0 = 0.2745$, $(M/R)_{\infty} = 0.125$
\begin{tabular}{cccccccc}
$z_A$ & $q^{\rm max}$ & $\bar M$ & $\bar J$ & $\bar \Omega$ &
    $\bar r_A$ & $\bar r_B$ & $\bar r_C$ \\
\tableline
  0.00 & 0.1119  & 0.260578 & 0.14820 & 0.055 & 0.000 & 2.549 & 4.680\\
  0.10 & 0.1141  & 0.260614 & 0.14830 & 0.051 & 0.472 & 2.658 & 4.722\\
  0.20 & 0.1237  & 0.260665 & 0.14908 & 0.045 & 0.964 & 2.909 & 4.824\\
  0.30 & 0.1380  & 0.260810 & 0.15227 & 0.038 & 1.518 & 3.291 & 5.061\\
\end{tabular}
\end{center}
\caption{}
\end{table}

\begin{table}[h]
\begin{center}
$n = 2$, $\bar M_0 = 0.495$, $(M/R)_{\infty} = 0.05 $
\begin{tabular}{cccccccc}
$z_A$ & $q^{\rm max}$ & $\bar M$ & $\bar J$ & $\bar \Omega$ &
    $\bar r_A$ & $\bar r_B$ & $\bar r_C$ \\
\tableline
  0.00 & 0.0381  & 0.48628 & 0.7204 & 0.0073 & 0.000 & 12.89 & 23.47\\
  0.10 & 0.0383  & 0.48635 & 0.7281 & 0.0068 & 2.382 & 13.49 & 23.82\\
  0.20 & 0.0389  & 0.48649 & 0.7508 & 0.0057 & 4.998 & 15.11 & 24.99\\
  0.30 & 0.0395  & 0.48672 & 0.7924 & 0.0045 & 8.192 & 17.78 & 27.30\\
\end{tabular}
\end{center}
\caption{}
\end{table}

\begin{table}[h]
\begin{center}
$n = 2$, $\bar M_0 = 0.52$, $(M/R)_{\infty} = 0.065$
\begin{tabular}{cccccccc}
$z_A$ & $q^{\rm max}$ & $\bar M$ & $\bar J$ & $\bar \Omega$ &
    $\bar r_A$ & $\bar r_B$ & $\bar r_C$ \\
\tableline
  0.00 & 0.0493  & 0.50929 & 0.7190 & 0.0095 & 0.000 & 10.70 & 19.52\\
  0.10 & 0.0497  & 0.50936 & 0.7255 & 0.0089 & 1.976 & 11.18 & 19.76\\
  0.20 & 0.0525  & 0.50950 & 0.7409 & 0.0077 & 4.069 & 12.30 & 20.34\\
  0.30 & 0.0574  & 0.50969 & 0.7679 & 0.0065 & 6.432 & 13.95 & 21.44\\
\end{tabular}
\end{center}
\caption{}
\end{table}

\end{appendix}


\begin{references}

\bibitem{tamt93} S. E. Thorsett, Z. Arzoumanian, M. M. McKinnon and
	J. H. Taylor, Astrophys. J. {\bf 405}, L29 (1993)

\bibitem{tw89} J. H. Taylor and J. M. Weisberg,
	Astrophys. J. {\bf 345}, 434 (1989)

\bibitem{a95} Z. Arzoumanian, PhD thesis, Princeton University (1995)

\bibitem{he84} I. Hachisu and Y. Eriguchi,
        Publ. Astron. Soc. Japan {\bf 36}, 239 (1984)

\bibitem{rs92} F. A. Rasio and S. L. Shapiro,
	Astrophys. J. {\bf 401}, 226 (1992); {\bf 432}, 242 (1994)

\bibitem{sno92} M. Shibata, T. Nakamura and K. Oohara,
        Prog. Theor. Phys. {\bf 88}, 1079 (1992)

\bibitem{zcm94} X. Zhuge, J. M. Centrella and S. L. W. McMillan,
        Phys.~Rev.~D {\bf 50}, 6247 (1994); {\bf 54}, 7261 (1996);

\bibitem{rjs95} M. Ruffert, H.-T. Janka and G. Sch\"afer,
        Astrophys. Sp. Sci. {\bf 231}, 423 (1995)

\bibitem{bdiww95} L. Blanchet, T. Damour, B. R. Iyer, C. M. Will
        and A. G. Wiseman, Phys.~Rev.~Lett. {\bf 74}, 3515 (1995) and
        references therein

\bibitem{s96} M. Shibata, Prog. Theor. Phys. {\bf 96}, 317 (1996);
	Phys. Rev. D. {\bf 55}, 6019 (1997)

\bibitem{ts97} K. Taniguchi and M. Shibata, 1997, to appear in 
	Phys. Rev. D. (gr-qc/9705027); M. Shibata and K. Taniguchi,
	1997, to appear in Phys. Rev. D. (gr-qc/9705028)


\bibitem{on96} K. Oohara and T. Nakamura, Lecture delivered at 
	Les Houches School ``Astrophysical Sources of Gravitational
     	Radiation'' (Les Houches, France, Sept. 26 - Oct. 6, 1995). To be
     	published in the Proceedings (eds. J.-A. Marck and J.-P. Lasota)
	(astro-ph/9606179)

\bibitem{l96} D. Lai, Phys.~Rev.~Lett. {\bf 76}, 4878 (1996)

\bibitem{lw96} D. Lai and A. D. Wiseman,
	Phys.~Rev.~D {\bf 54}, 3958 (1996)

\bibitem{lrs97} J. C. Lombardi, F. A. Rasio and S. L. Shapiro, 1997, 
	Phys. Rev. D, in press (astro-ph/9705218)

\bibitem{wm95} J. R. Wilson and G. J. Mathews,
        Phys.~Rev.~Lett. {\bf 75}, 4161 (1995);
	J. R. Wilson, G. J. Mathews and P. Marronetti,
	Phys.~Rev.~D {\bf 54}, 1317 (1996) (WMM)

\bibitem{lrs93} D. Lai, F. A. Rasio and S. L. Shapiro,
	Astrophys. J. Suppl. {88}, 205 (1993)

\bibitem{w97} A. D. Wiseman, Phys. Rev. Lett. {\bf 79}, 1189 (1997)

\bibitem{bh97} P. R. Brady and S. A. Hughes, 
	Phys. Rev. Lett. {\bf 79}, 1186 (1997) 

\bibitem{f97} \'E. \'E. Flanagan, 1997, submitted (gr-qc/9706045)

\bibitem{t97} K. S. Thorne, 1997, submitted (gr-qc/9706057)

\bibitem{eh96} D. M. Eardley and E. W. Hirschmann,
	preprint NSF-ITP-95-165 (1995) (astro-ph/9601019)

\bibitem{bcsst97a} T. W. Baumgarte, G. B. Cook, M. A. Scheel, 
	S. L. Shapiro and S. A. Teukolsky, 
	Phys. Rev. Lett. {\bf 79}, 1182 (1997)

\bibitem{bcsst97b} T. W. Baumgarte, G. B. Cook, M. A. Scheel,
	S. L. Shapiro and S. A. Teukolsky, 1997, submitted
	(gr-qc/9705023)

\bibitem{bc92} L. Bildsten and C. Cutler, 
	Astrophys. J. {\bf 400}, 175 (1992)

\bibitem{k92} C. S. Kochanek, Astrophys. J. {\bf 398}, 234 (1992)

\bibitem{wm89} J. R. Wilson and G. J. Mathews, in
        {\em Frontiers in Numerical Relativity}, edited by C. R. Evans,
        L. S. Finn, and D. W. Hobill (Cambridge University Press, 
        Cambridge, England, 1989), pp. 306;

\bibitem{cst96} G. B. Cook, S. L. Shapiro and S. A.Teukolsky,
        Phys.~Rev.~D {\bf 53}, 5533 (1996)

\bibitem{adm62} R. Arnowitt, S. Deser and C. W. Misner, in 
	{\em Gravitation}, edited by L. Witten (Wiley, New York, 1962)

\bibitem{y72} J. W. York, Jr., Phys. Rev. Lett. {\bf 28}, 1082 (1972);
	Niall \'O Murchadha and J. W. York, Jr., Phys. Rev. D, {\bf 10},
	428 (1974)

\bibitem{fn1} Note that we could have chosen any metric for which 
	$\gamma^{-1/3}\gamma_{ij}$ is independent of time. Choosing a
	conformally flat metric leads to familiar flat-space
	differential operators.

\bibitem{by80} J. M. Bowen and  J. W. York, Jr.,
	Phys.~Rev.~D {\bf 21}, 2047 (1980)

\bibitem{b82} J. M. Bowen, Gen. Rel. Grav. {\bf 14}, 1183 (1982)

\bibitem{lppt75} A. P. Lightman, W. H. Press, R. H. Price and 
	S. A. Teukolsky, {\em Problem Book in Relativity and Gravitation}
	(Princeton University Press, Princeton, New Jersey, 1975),
	problem 16.17

\bibitem{fn2} We will loosely refer to $q$ as the density, even though 
	the density is really $\kappa^{-n}q^n$.

\bibitem{h86} I. Hachisu, Astrophys. J. Suppl. {\bf 61}, 479 (1986)

\bibitem{keh89} H. Komatsu, Y. Eriguchi and I. Hachisu,
	Mon. Not. R. Astron. Soc. {\bf 237}, 355 (1989)

\bibitem{cst92} G. B. Cook, S. L. Shapiro and S. A.Teukolsky,
	Astrophys. J. {\bf 398}, 203 (1992); Astrophys. J. {\bf 422},
	227 (1994)

\bibitem{ptvf92} W. H. Press, S. A. Teukolsky,  W. T. Vetterling and
	B. P.Flannery, {\em Numerical Recipes in Fortran, Second Edition}
	(Cambridge University Press, Cambridge, 1992)

\bibitem{fn3} Note that we limit this search to the $z$-axis, whereas 
	in general the maximum density may occur elsewhere inside the 
	star.  However, this is not a limitation of our method and also 
	we have found that in all our simulations the maximum density 
	did indeed occur on the $z$-axis.

\bibitem{pb95} M. Parashar and J. C. Brown, in {\em Proceedings of 
	the International Conference for High Performance Computing}, 
	eds. S. Sahni, V. K. Prasanna and V. P. Bhatkar
	(Tata McGraw-Hill Publishing Company Ltd, 1995), also 
	{\tt www.ticam.utexas.edu/$\sim$parashar/public\_html/DAGH/}

\bibitem{h70} J. B. Hartle, Astrophys. J. {\bf 161}, 111 (1970)

\bibitem{b72} J. M. Bardeen, in {\em Black Holes}, ed. C. DeWitt
	(New York: Gordon \& Breach, 1972)

\bibitem{c70} S. Chandrasekhar, Astrophys. J. {\bf 161}, 561 (1970)

\bibitem{fis88} J. L. Friedman, J. R. Ipser and R. D. Sorkin, 
	Astrophys. J. {\bf 325}, 722 (1988)

\bibitem{kww93} L. E. Kidder, C. M. Will and A. G. Wiseman, 
	Phys. Rev. D {\bf 47}, 3281 (1993)

\bibitem{h86} I. Hachisu, Astrophys. J. Suppl. {\bf 62}, 461 (1986)

\bibitem{rs94} F. A. Rasio and S. L. Shapiro, Astrophys. J. {\bf 432},
	242 (1994)

\bibitem{st83} S. L. Shapiro and S. A. Teukolsky, {\em Black Holes,
	White Dwarfs, and Neutron Stars} (New York, Wiley, 1983)
	 


\end{references}
\end{document}